\documentclass[11pt]{article}

\usepackage{amsmath}
\usepackage{graphicx}
\renewcommand{\baselinestretch}{1.5}
\begin{document}
\title{Assessing the value of cooperation in Wikipedia}%
\author{Dennis M. Wilkinson and Bernardo A. Huberman \\ HP Labs, Palo Alto, CA 94304}%
\maketitle

\begin{abstract}
\noindent Since its inception six years ago, the online
encyclopedia Wikipedia has accumulated 6.40 million
articles and 250 million edits, contributed in a
predominantly undirected and haphazard fashion by
5.77 million unvetted volunteers. Despite the apparent lack of order, 
the 50 million edits by 4.8 million
contributors to the 1.5 million articles in the
English-language Wikipedia follow strong certain overall
regularities. We show that the accretion of edits to an article is described by 
a simple stochastic mechanism, resulting in a heavy tail of highly visible articles with a large number of edits. 
We also demonstrate a crucial correlation
between article quality and number of edits, which
validates Wikipedia as a successful collaborative
effort.
\end{abstract}

\newpage

\subsection*{Introduction}\label{s:intro}
The online encyclopedia Wikipedia\footnote{http://wikipedia.org} is an impressive
example of a global collective intelligence at
work. Since its inception in January 2001, Wikipedia has grown to
encompass 6.40 million articles in 250
languages generated from
236 million edits by 5.77 million
contributors\footnote{http://meta.wikimedia.org/wiki/List\_of\_Wikipedias}, as of this writing. Its growth has been exponential in key metrics such as number of editors and number of articles \cite{Voss}.
That the content of
Wikipedia is deemed useful and relevant by the user
community at large is confirmed by its current position as 11th most visited site on the Internet\footnote{http://www.alexa.com/}, serving an average of 16536 requests per second\footnote{http://hemlock.knams.wikimedia.org/$\sim$leon/stats/reqstats/reqstats-monthly.png}. 

Since Wikipedia lets any user modify any article or create new articles, it virtually eliminates the barrier to contribution. This scheme paves the way for rapid expansion, but at uncertain cost to the article quality and value. It is of interest to understand Wikipedia's growth and to assess the quality and value of its articles, both to evaluate Wikipedia as a cooperative process and because of its great popularity. A number of recent studies have focused on these goals.

Wikipedia has been studied in the context of network dynamics \cite{Zlotic}, with the addition of new articles described by a time-dependent acceleration mechanism \cite{Smith} or a somewhat puzzling preferential attachment model \cite{Capocci}. Other work has examined the evolution of editors' roles and contributions in Wikipedia's development \cite{Kittur}. A power law relation for a fraction of the distribution of edits per article has been observed \cite{Buriol,Voss}, but no mechanism was proposed. While all this work contributes to the understanding of Wikipedia as a complex system, it does not provide insight into its development at the level of individual articles. 

A number of methods for automatic assessment of the quality of Wikipedia's articles have also been proposed. In \cite{Lih}, the number of edits and unique editors to an article were suggested as metrics for quality, but no justification was provided. Other characteristics such as factual accuracy \cite{Nature,Brit,Nature_rebuttal}, credibility \cite{Chesney}, revert times \cite{Viegas}, and formality of language \cite{Emigh} have been used to assess small samples of Wikipedia's articles and in some cases compare them to articles of traditional encyclopedias. It is doubtful that encyclopedia quality can be assessed using a single metric (e.g. \cite{Crawford}), but complex combinations of metrics \cite{Stviliamain} depend on rather arbitrary parameter choices. A crucial element lacking from previous tests of metrics of article quality is any consideration of article popularity or relevance, which can clearly be expected to affect metrics such as number of edits, number of links, article length, number of images, and many others.

In this paper we first show that Wikipedia articles accrete edits according to a simple stochastic mechanism resulting in a population of disproportionally highly-edited articles. We then demonstrate a strong correlation between number of edits and article quality. Topics of particular interest or relevance are thus naturally brought to the forefront of quality. This is significant because Wikipedia is frequently used as a source of information, and because other large collaborative efforts such as software development \cite{Brooks}, industrial design \cite{Allen} and cooperative problem solving \cite{Clearwater} are known to produce ambiguous results as the size of the project increases. 

At the heart of the evolution of a Wikipedia articles is the simple rule edits beget edits. That is, the number of new edits to a given article at a given time is a randomly varying percentage of the total number of previous edits. This process produces a lognormal distribution in the number of edits per article for articles created during a particular time slice, where the distribution parameters $\mu$ and $\sigma^2$ depend linearly on the age of the time slice. The lognormal distribution implies than while most articles accrete only a small number of edits, there is a significant population of articles with a disproportionally large number of edits. Additionally, the increase in $\mu$ with article age implies that articles continue to accrete edits and do not reach a steady state.

To examine the correlation between edit volume and article quality, we compared the average number of edits and contributors on ``featured'' articles, selected by the Wikipedia community as ``the best articles in Wikipedia,'' to the corresponding averages for other articles. The results show a strong correlation between number of edits, number of distinct editors, and article quality. In making this comparison, it is crucially important to control for the article visibility or relevance, since featured articles tend to deal eith more popular subjects. Article age must also be taken into consideration, since on average older articles have more edits.  Care was taken to control for these variables.

\subsection*{Article growth}
\label{s:mech}
While individual users exhibit highly variable editing activities, the overall pattern of how articles accrete edits is well-described by the simple stochastic mechanism described as follows. 

Consider the number of new edits $\Delta n(t)$ to an article made between time $t$ and time $t+dt$, an interval of perhaps several hours. Of course, complicated fluctuations in human behavior and activity cause this number to vary in a random way, but we claim that $\Delta n(t)$ is on average proportional to the total number of previous edits. This is expressed mathematically as 
\[
\Delta n(t) = [a+\xi(t)] n(t),
\] 
where $n(t)$ is the total number of edits to a given article up until time $t$, $a$ is a constant (average) rate of edit accretion, and $\xi(t)$ is mean-zero random term accounting for fluctuations. The total number of edits at time $t+dt$ is thus given by 
\begin{equation}\label{e:main}
	n(t+dt)= n(t) + \Delta n(t) = [1 + (a + \xi(t))] n(t).
\end{equation}

Because of the random nature of human activity embodied by $\xi(t)$, the number of edits to a given article at a given time can be predicted only within a range of values specified by a probability distribution. Previous work on similar processes, such as World Wide Web traffic \cite{Huberman} and many others (e.g., \cite{Ross}), has shown that the distribution resulting from equation (\ref{e:main}) is lognormal\footnote{In equation \ref{e:main}, the noise terms at different $t$ are assumed to be uncorrelated. In fact, as one might expect, the percentage increase in edits does demonstrate a small positive autocorrelation over periods of less than 20 to 30 days. Since the autocorrelation length is finite, however, the central limit theorem may still be applied to obtain a log-normal distribution; the difference is that the rate parameter $a$ must be modified to account for the autocorrelation \cite{Berk}. Because the modification is small, for the sake of simplicity, we do not include it here.} and given by
\begin{equation}\label{e:logn}
	P[n(t)] = \frac{1}{n\sqrt{2\pi}\sqrt{s^2t}}\exp{\left[-\frac{(\log n - at)^2}{2(s^2t)}\right]},
\end{equation}
where $s^2$ is the variance of the $\xi(t)$. This equation shows that the distribution parameters $\mu = at$ and $\sigma^2 = s^2t$ are linearly related to the age $t$ of the article. $\mu$ and $\sigma^2$ represent the mean and variance, respectively, of the log of the data, and are thus related to but not equal to the distribution mean and variance. In practice, we considered articles created during a time slice of average age $t$ in order to obtain enough data points to constitute a distribution. Provided the time slice is not too long, editing within the slice does not corrupt the distribution much.

\begin{figure}[t!] 
 \renewcommand{\baselinestretch}{1}
\centering
    \includegraphics[width=2.25in]{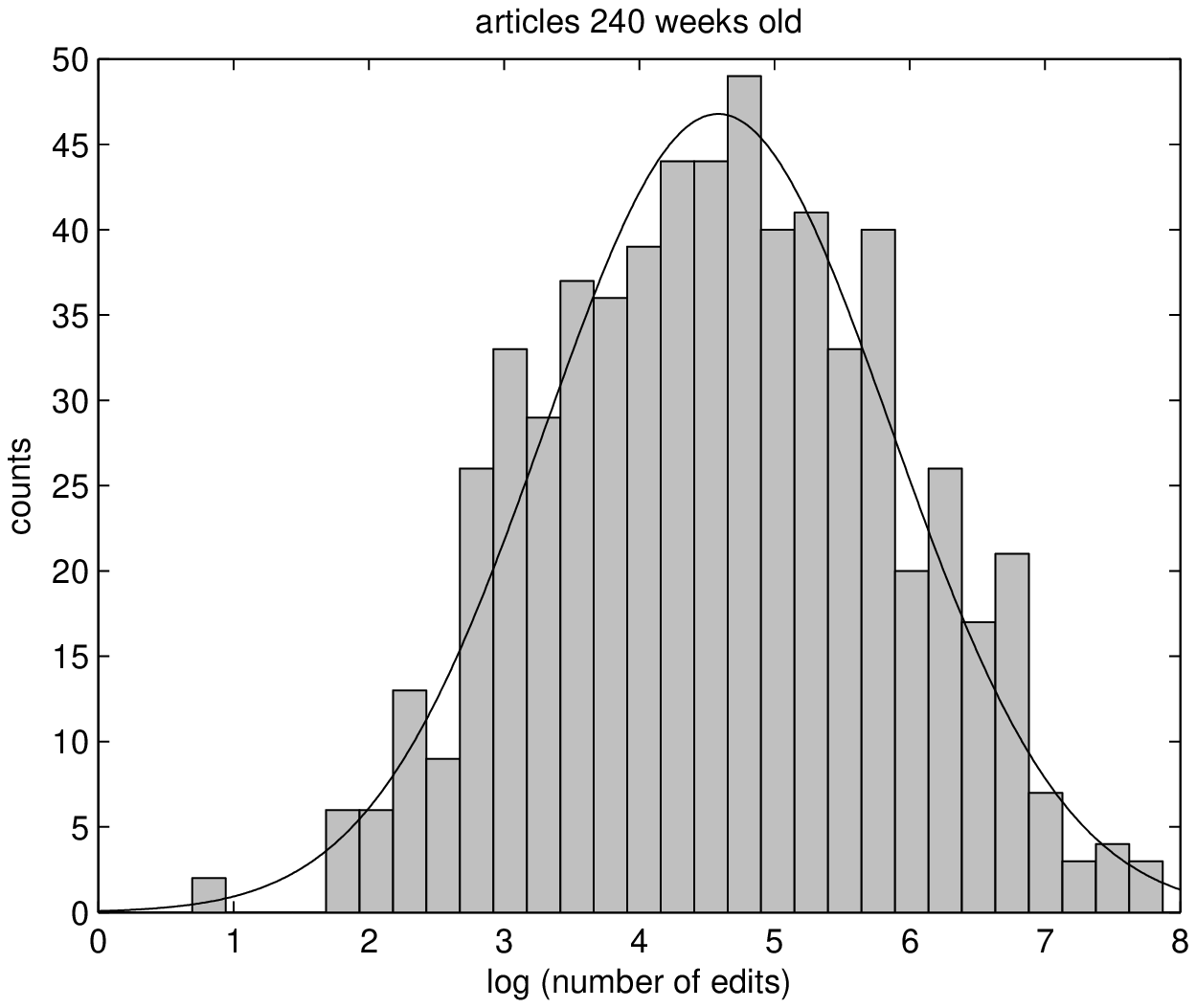}
    \includegraphics[width=2.25in]{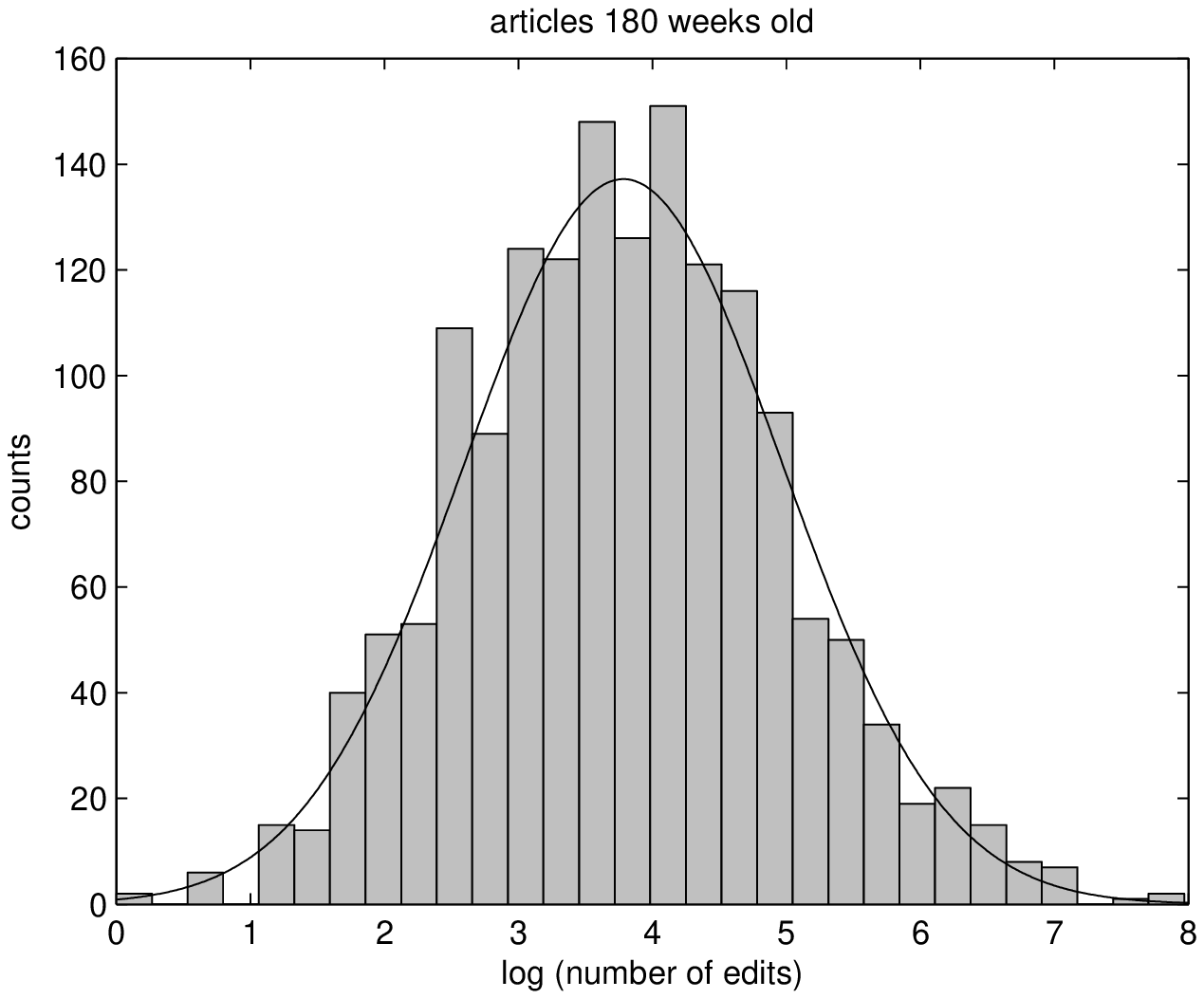}
    \includegraphics[width=2.25in]{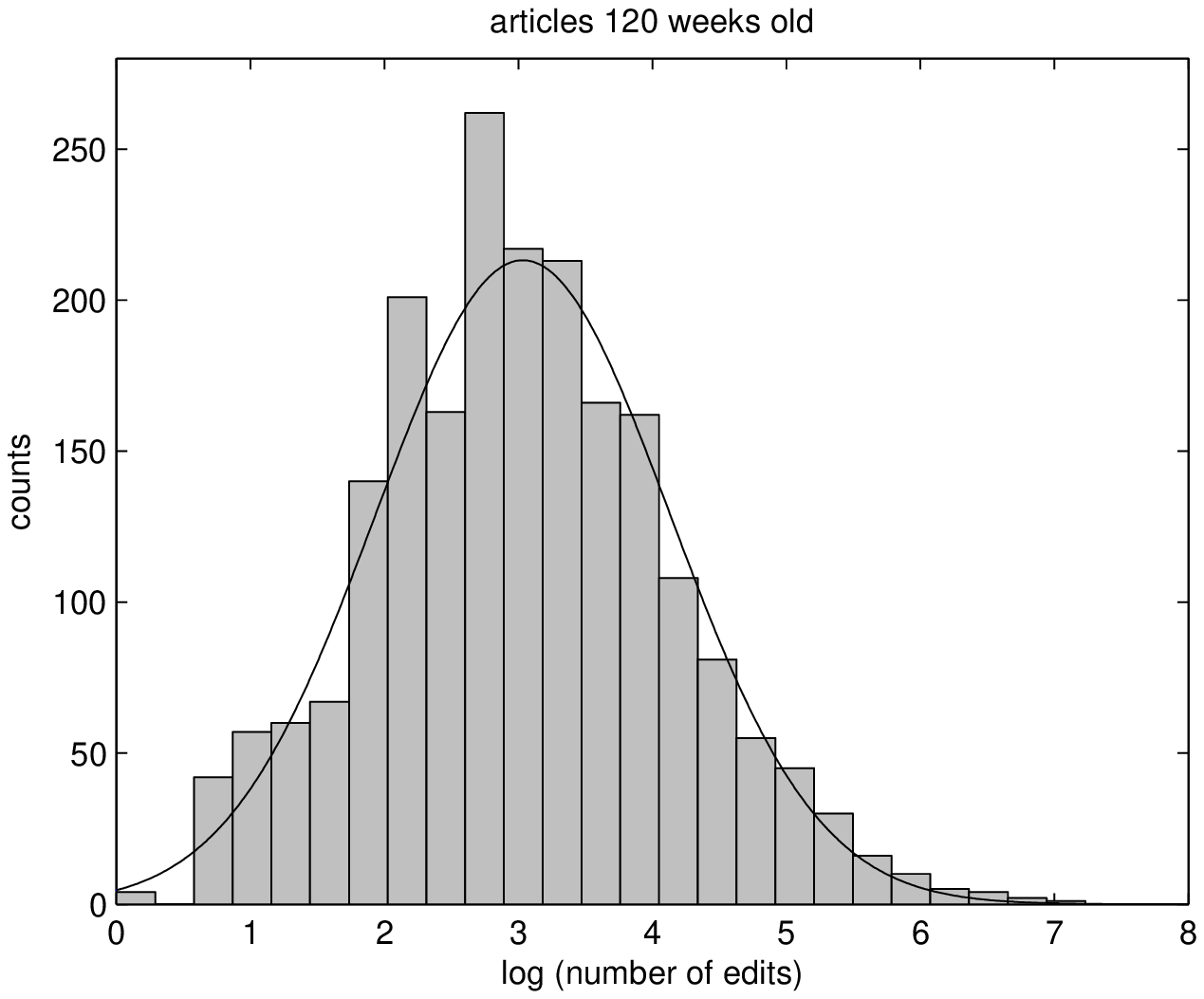}
    \caption{ \small Distributions of the logarithm of the number of edits per article for articles of ages $t$ = 240, 180, and 120 weeks. Because the distribution of the counts is lognormal, the logarithm of the counts should be normally distributed, and the best fit normal curve is included for comparison. Note how the distribution mean increases with age, as expected, while the number of counts per week increases (due to the overall growth of Wikipedia).}
    \label{f:dists}
\end{figure}

\begin{figure}[ht!] 
 \renewcommand{\baselinestretch}{1}
\centering
    \includegraphics[width=2.5in]{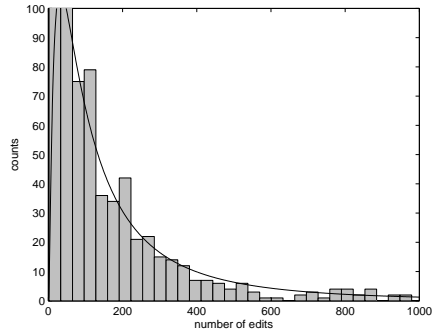}
    \caption{ \small Lognormal distribution for the number of edits per article for articles of age $t=240$ weeks. The plot was truncated at the high end of both axes for readability; in fact, there are articles in this time slice with many thousands of edits. The best fit lognormal curve is included for comparison.}
    \label{f:logndist}
\end{figure}

Equation (\ref{e:logn}) is verified by a study of the 50.0 million edits made by the 4.79 million non-robot contributors to the 1.48 million articles of the English-language Wikipedia between its inception in January 2001 and November 2, 2006. A statistical test of all time slices yields a $p$-value of greater than 0.5 for 50.9 \% of the 3688 relevant time slices for the lognormal distribution (further details on the test and the data are provided in the appendix). The shape of the distribution of edits for articles in various time slices is best displayed using a histogram of the logarithm of the edit counts, which follows a normal distribution with mean $\mu(t)$ and variance $\sigma^2(t)$, as shown in figure \ref{f:dists}. The actual lognormal distribution for one time slice, showing only a portion of the tail, is pictured in figure \ref{f:logndist}.
\begin{figure}[h!] 
 \renewcommand{\baselinestretch}{1}
\centering
    \includegraphics[width=3.5in]{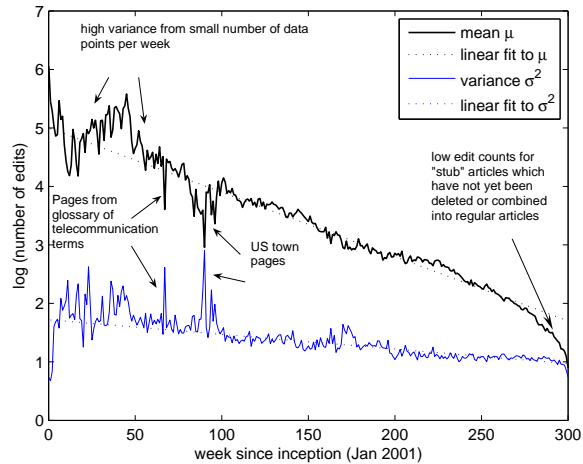}
    \caption{ \small Evolution of the mean $\mu$ and variance $\sigma^2$ of the lognormal distribution of edits per article. The $x$-axis represents the week in which articles were created, and the $y$-axis the mean $\mu$ or variance $\sigma^2$ of the log of the number of edits to articles created during that week. The deviations from the pattern include two periods in which a large number of rather trivial articles with low edit counts were created at once, and the recent data containing a large number of short ``stub'' articles which have yet to be combined into regular articles or deleted.}
    \label{f:musig}
\end{figure}

The variation of the distribution parameters $\mu$ and $\sigma^2$ with age is demonstrated in figure \ref{f:musig}. The linear dependence is highlighted by the fitted curve. Anomalous time slices which do not fit the overall trend are noted in the figure as well. Because of a single editor's activity, these slices contain an unusually high number of articles with low edit counts.

The lognormal distribution has a heavy tail at the high end, implying that a small number of articles accrete a disproportionally large number of edits. As we show below, edits correspond on average to an increase in article quality. The multiplicative mechanism of edit accretion thus creates a small body of high quality articles. These high quality articles deal with topics of high visibility or relevance, while the vast majority of Wikipedia articles are relatively infrequently edited and have far lower visibility.

Since each time slice follows a lognormal distribution, the overall distribution of edits per article is a mixture over time of lognormals with linearly varying $\mu$ and $\sigma^2$, multiplied by a factor accounting for the overall growth of Wikipedia. This integral is not analytic for the time frame of our data, but by numerical integration can be shown to retain a lognormal character\footnote{In the long-time limit, the integral tends towards a power law if overall growth is exponential \cite{Huberman}. However, the time frame of our data set is not the long time limit, in contrast to the findings of \cite{Buriol} who report a power law fit for the overall distribution of edits per article.}.

\subsection*{Edits and article quality}
As discussed in the introduction, it is of considerable interest to determine whether extensive editing by a large number of diverse contributors increases article quality. To test for a correlation between editing and article quality, we compared the number of edits and contributors to the 1211 ``featured'' articles to the corresponding numbers for other articles. As mentioned above, featured articles are those selected by the Wikipedia community as ``the best articles in Wikipedia'' according to criteria such as accuracy, neutrality, completeness, and style\footnote{http://en.wikipedia.org/wiki/Wikipedia:Featured\_articles}. Featured articles which do not continue to uphold these high standards are demoted.

Care must be taken when comparing edit volumes on different populations of articles. First of all, the relevance or visibility of an article's topic must be taken into account. Secondly, the age of the article also affects its number of edits, as demonstrated by figure \ref{f:musig}. 
\begin{figure}[htb]
 \renewcommand{\baselinestretch}{1}
\centering
    \includegraphics[width=2.5in]{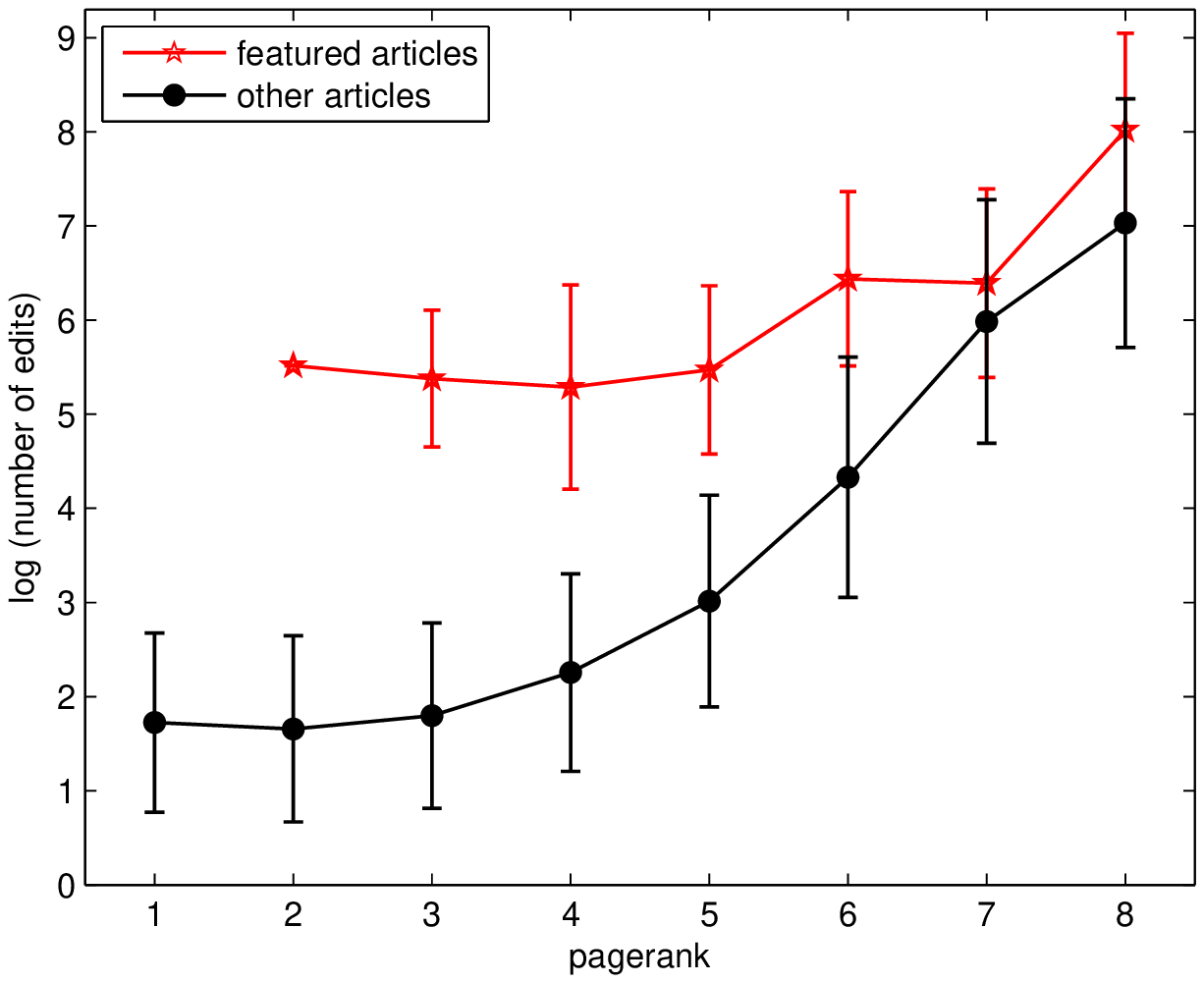}
    \includegraphics[width=2.5in]{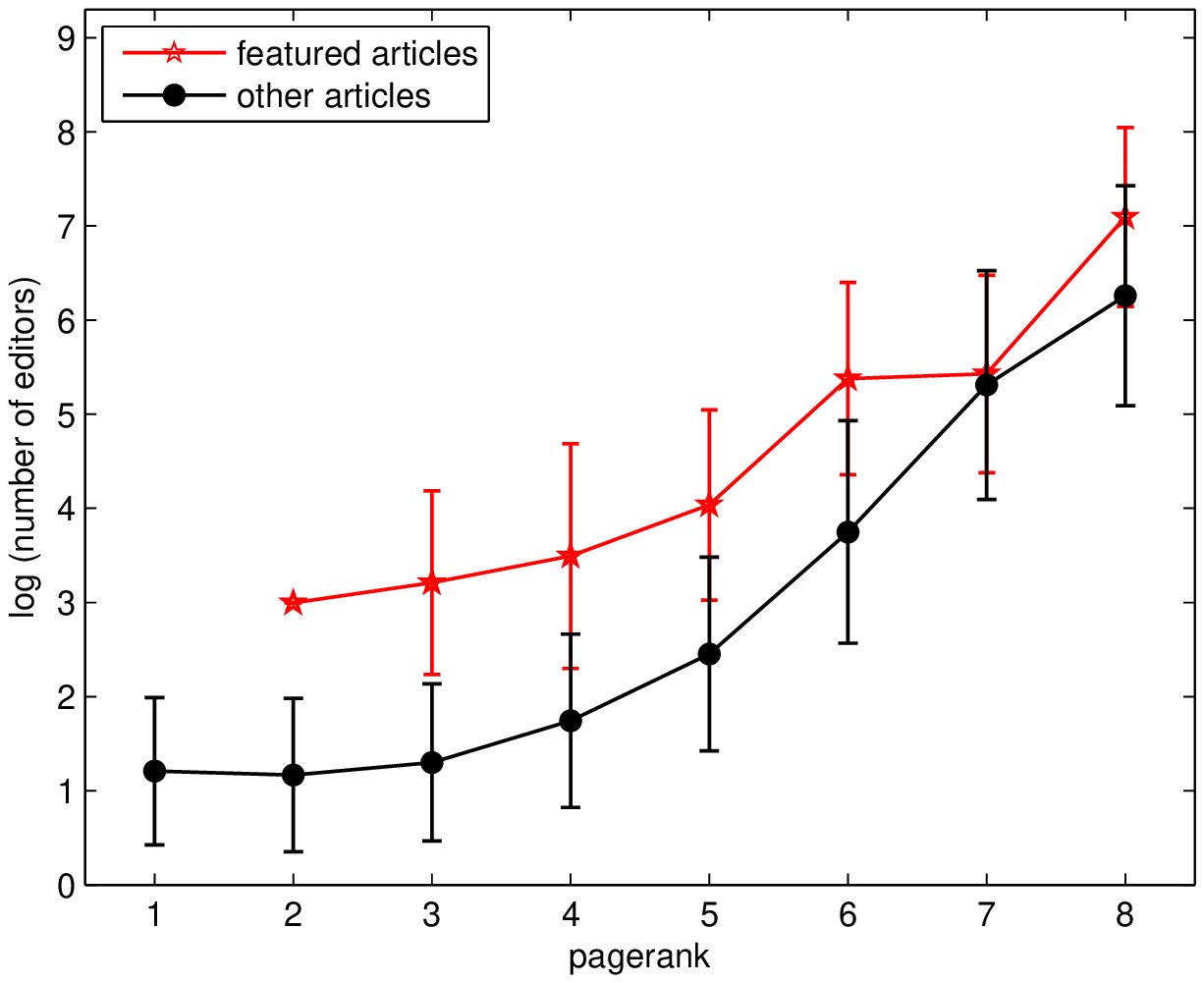}
    \caption{\small Average and standard deviation (error bars) for the log of the number of edits, at left, and number of distinct editors, at right, per article. The articles are grouped by pagerank for reasons of relevance and visibility as discussed in the text.}
     \label{f:eds}
\end{figure}

To control for article visibility or relevance, we grouped articles by their Google pagerank before comparing them. We also controlled for the added visibility some featured articles obtain when they appear on the Wikipedia front page by discounting the edits made during this period. These results, which do not yet account for article age, are shown in figure \ref{f:eds}. For all pageranks except 7, a strong correlation between edits or editors and quality is seen. The anomalous behavior at pagerank 7 disappears when age is accounted for, below\footnote{Many articles of high interest or relevance were among the earliest created, so there is a predominance of high pageranks among the oldest Wikipedia articles.}.

\begin{figure}[htb]
 	\renewcommand{\baselinestretch}{1}
	\centering
  \includegraphics[width=3in]{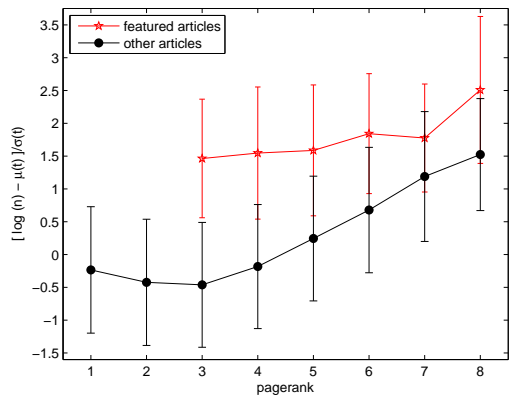}
  \caption{\small Average and standard deviation (error bars) of the age-normalized measure of edit volume (equation \ref{e:norm}), grouped by pagerank.}
  \label{f:norm}
\end{figure}

To control for article age, we normalized (the logarithm of) the number of edits to an article of age $t$ by the mean and variance for all articles of that age, as previously computed (figure \ref{f:musig}). For a given article $A$ of age $t$ having undergone $n(A)$ edits, we thus computed the difference between $\log[n(A)]$ and the average $\mu(t)$ of the logarithm of the number of edits of age $t$, in units of $\sigma(t)$: 
\begin{equation}\label{e:norm}
x(A) = \frac{\log n(A)  - \mu(t)}{\sigma(t)}.
\end{equation}
The featured and nonfeatured populations were then compared using this metric, with the results shown in figure \ref{f:norm}.

The plots of figures \ref{f:eds} and \ref{f:norm} show a strong correlation between number of edits, number of distinct editors, and article quality in Wikipedia. The heavy tail of articles with disproportionally high edit counts discussed previously thus represents a collection of predominantly high quality articles. It is also interesting that pagerank, reportedly a logarithmic scale, is more or less linearly related to the number of edits or editors of an article\footnote{While the figure shows the average of the log of number of edits, a similar plot for the log of the average produces a similar, nearly linear relation.}.

As to the question of causality between edits and quality, recall that articles always continue to accrete edits and evolve instead of reaching a steady state, as we showed. Resolving causality in an aggregate sense is thus most likely impossible. Indeed, the development of an article is a highly complex process \cite{Stviliarecent} and both directions of causality between editing and quality are likely to play a role.

\subsection*{Conclusion}
We have shown that although Wikipedia is a complex system in which of millions of diverse editors collaborate in an
unscheduled and virtually uncontrolled\footnote{In fact, a group of dedicated ``administrative users'' have provided ideological guidance, supervision in the worst cases or dispute or vandalism, and a relatively large number of contributions to Wikipedia. However, the process of editing remains almost completely unsupervised.} fashion, editing follows a very simple overall pattern. This pattern implies that a small number of articles, corresponding to topics of high relevance or visibility, accrete a disproportionately large number of edits.

And, while large collaborations have been shown to fail in many contexts, Wikipedia article quality continues to increase, on average, as the number of collaborators and the number of edits increases. Thus, topics of high interest or relevance are naturally brought to the forefront of visibility and quality. 

\vspace{0.3in}

{\bf \noindent Acknowledgments:} We thank Travis Kriplean for his work in helping process the data set and Yuri Karaban for his Perl module.

\vspace{0.2in}

\subsection*{Appendix: Methods}\label{s:methods}
The raw data for our study were all 55.3 million edits to the English-language Wikipedia made between Wikipedia's inception in January 2001 and November 2, 2006. This data included username or url, page title, and timestamp\footnote{The data is publicly available at http://meta.wikimedia.org/wiki/Data\_dumps}. From the raw data, we eliminated redirect and disambiguation pages, which are articles with no content that merely point to other articles, and edits made by robots. Redirects and disambiguation pages were identified using simple text analysis. Robot edits were identified using the list of registered Wikipedia robots\footnote{http://en.wikipedia.org/wiki/Wikipedia:Registered$\_$bots}, and by identifying edits made by a single user in improbably quick succession. This process eliminated 5.23 million edits, or 9.5 \% of the original 55.3 million. 

A small percentage of articles were not used because of technical difficulties in the title caused by rare foreign characters. Google pageranks were obtained by querying Google\footnote{Yuri Karaban's PERL module was very helpful for this: http://search.cpan.org/$\sim$ykar/WWW-Google-PageRank-0.12/}. Some recent articles did not yet have a pagerank and so pagerank zero articles were not included in the analysis. 

To test the lognormal fit and obtain the quoted $p$-value, we applied a typical $\chi^2$ fitting procedure to each time slice using the likelihood ratio statistic \cite{Rice}. In this test, the time slice length was variable because of the overall growth of Wikipedia; more recent articles were grouped into smaller slices because otherwise the distribution was skewed by edits made within the slice. In practice, we chose time slices long enough to contain 400 articles. The expected distribution for each slice was calculated using the slice's sample mean and variance, and the data was grouped into bins whose width was the minimum required to make the expected count greater than 8. Of course, slight variations in the quoted $p$-value, on the order of several percent, were obtained by varying the time slice length and bin size.

\newpage

\footnotesize

\bibliography{wik_arxiv}

\bibliographystyle{plain}

\end{document}